# The Democratic Paradox in Large Language Models' Underestimation of Press Freedom

Isabella Loaiza[1*†], Roberto Vestrelli[1,2†],
Andrea Fronzetti Colladon[3], Roberto Rigobon[1]

[1*]Sloan School of Management, Massachusetts Institute of Technology, .
[2]Department of Engineering, University of Perugia, .
[3]Department of Civil, Computer Science and Aeronautical Technologies Engineering, Roma Tre University, .

*Corresponding author(s). E-mail(s): isal@mit.edu; †These authors contributed equally to this work.

**Abstract**

As Large Language Models (LLMs) increasingly mediate global information access for millions of users worldwide, their alignment and biases have the potential to shape public understanding and trust in fundamental democratic institutions, such as press freedom. In this study, we uncover three systematic distortions in the way six popular LLMs evaluate press freedom in 180 countries compared to expert assessments of the World Press Freedom Index (WPFI). The six LLMs exhibit a negative misalignment, consistently underestimating press freedom, with individual models rating between 71% to 93% of countries as less free. We also identify a paradoxical pattern we term differential misalignment: LLMs disproportionately underestimate press freedom in countries where it is strongest. Additionally, five of the six LLMs exhibit positive home bias, rating their home countries' press freedoms more favorably than would be expected given their negative misalignment with the human benchmark. In some cases, LLMs rate their home countries between 7% to 260% more positively than expected. If LLMs are set to become the next search engines and some of the most important cultural tools of our time, they must ensure accurate representations of the state of our human and civic rights globally.

**Keywords:** Large Language Models, Press Freedom, Differential Misalignment, Home Bias, Media Trust

# 1 Introduction

Democratic governance depends fundamentally on citizens' trust in institutions. This trust connects individuals to the systems designed to serve them, enhancing both legitimacy and governmental effectiveness [1]. The press, serving as democracy's information infrastructure, enables informed civic participation and ensures governmental accountability. Thus, confidence in the media and its integrity represents a particularly critical component of this institutional ecosystem.

Trust in the media is associated with both trust in democracy itself [2] and measurable voting behavior. Citizens with diminished confidence in the press rely more heavily on partisan ideology in voting decisions [3], support populist candidates more frequently, and abstain from voting at higher rates [4]. Trust in the media, however, has declined in several countries, with the United States experiencing one of the most significant decreases among developed nations [5]. This erosion of confidence in the press has been driven by multiple factors, including the increase of internet-based information sources [6], exposure to partisan news content [7], perceptions that the media can be subject to commercial and governmental interests [8], and the rise of social media platforms [9].

Due to its critical role in reshaping the digital information ecosystem, researchers have sought to understand its impact on media trust and other democratic institutions, finding both positive and negative effects [10]. On the one hand, by enabling distributed content sharing and allowing human connection across distances, social media has enabled self-organized political behavior and civic stewardship [11]. On the other hand, through algorithmic content curation, social media has contributed to the creation of echo chambers that reinforce partisan beliefs [12] and the broader spread of false news compared to truth [13].

While using social media to consume news does not necessarily correlate with media distrust, individuals who rely on it exclusively demonstrate decreased confidence in the press [9]. Furthermore, a recent systematic review examining the relationship between digital media and democracy concluded that while social media fosters participation and civic action, it has predominantly had detrimental effects on institutional trust [10]. These findings raise critical questions about how other algorithmically mediated technologies, such as Large Language Models (LLMs), might influence public perception of the press and its integrity.

LLMs represent a fundamentally new form of algorithmic mediation in information consumption, rapidly reshaping how millions of users access information worldwide. As these systems become key cultural [14] and geopolitical [15, 16] tools, they must ensure accurate representations of democratic institutions, like the press and human and civic rights. Unlike social media's distributed nature, LLMs resemble traditional broadcasters in that they are centralized, yet they share the characteristic of algorithmic content mediation. Additionally, LLMs possess unique features that set them apart from social media, such as their sycophantic behavior [17]. Given these features, how might LLMs reshape users' perceptions of the press in ways that are distinct from social media?

Extensive research documents LLM biases toward demographic groups, but less attention has been paid to understanding how LLMs portray democratic institutions. Previous research has documented cultural [18], political [19–22], and geographical [23] biases in LLMs. There is also evidence of social biases against some social groups based on gender [24, 25] or in support of one's group, commonly known as in-group bias [26]. At the international level, recent studies suggest that LLM responses can reflect national alignments, particularly when prompted in languages other than English. Durmus et al. [27] found that LLMs, when prompted in non-English languages, often disproportionately reflect values prevalent in the United States and select European and South American populations while under-representing alternative viewpoints.

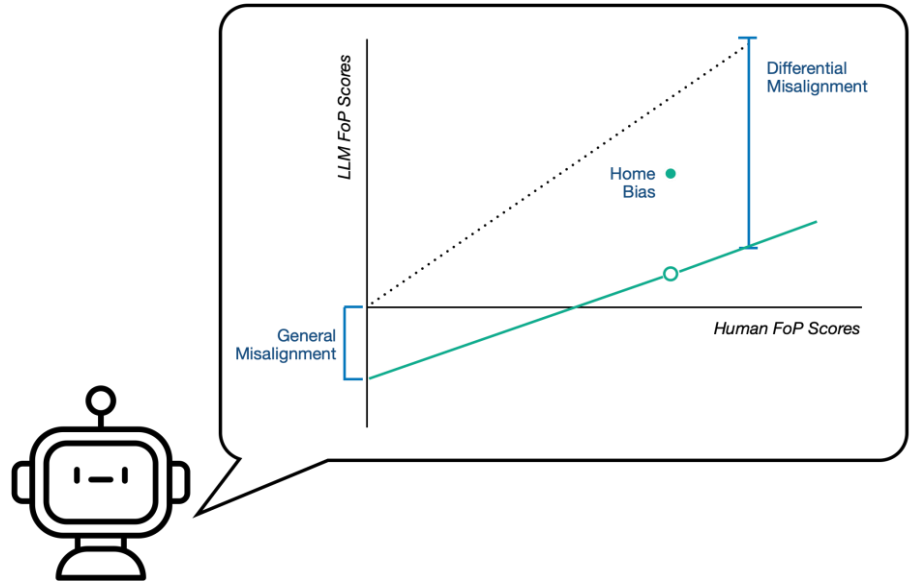

**Fig. 1 Factors that shape LLM evaluations of press freedom.** This figure illustrates three patterns identified in our study that contribute to LLM misevaluations of press freedom compared to expert assessments from the World Press Freedom Index (WPFI). The schematic representation shows LLM scores (y-axis) plotted against human expert scores (x-axis), where the dotted 45-degree line represents perfect alignment between LLM and human evaluations. **General misalignment:** LLMs consistently rate press freedom more critically than the WPFI benchmark (over 71% of the time for the least misaligned LLM), illustrated by the negative intercept of the model's score line (in green). **Differential misalignment:** The six LLMs exhibit varying degrees of misalignment, being more critical of countries with higher press freedom and less critical of those with restrictive regimes. This is represented by the LLM score line having a smaller slope than the 45-degree reference line. **Positive home bias:** Five of the six LLMs rate press freedom in their home countries more favorably than in other countries relative to their ratings of countries with similar levels of press freedom and their general misalignment, shown as the hollow circle that falls above the expected LLM trend line (green) but still below the 45-degree line.

Conversely, Zhou and Zhang [28] found that when prompted in Chinese, large language models produce noticeably different and more favorable portrayals of China compared to prompts in English.

For this reason, in this paper, we set out to study how LLMs portray a critical institution for democratic governance: freedom of the press. We examine how six widely used large language models assess press freedom across 180 countries. In this study, we evaluated ChatGPT, Gemini, DeepSeek, Qwen, Mistral, and Falcon using multiple-choice and open-ended prompts, comparing their scores against human expert assessments from the World Press Freedom Index (WPFI) published by RSF. Our analysis reveals three forms of systematic misalignment between LLM and expert evaluations of press freedom.

Our results reveal that the LLMs in our study have three different types of misalignment when evaluating press freedom. First, the six LLMs exhibit a generalized misalignment, characterized by consistently under-rating press freedom compared to the human benchmark. Second, the six models show evidence of what we call differential misalignment. This paradoxical differential misalignment entails that the magnitude of LLMs' misalignment is associated with countries' press freedom levels, leading LLMs to disproportionately score freer countries more negatively. Third, we find that five of the six language models exhibit a specific form of differential misalignment that we call home bias. We define home bias as an LLM's tendency to score its home country's press freedom more favorably than expected, given the model's generalized and differential misalignment with the human benchmark. A positive home bias entails rating home countries more favorably relative to the other two misalignment, but not necessarily above the human benchmark in absolute terms.

With this work, we contribute to the growing body of literature that seeks to deepen our understanding of the impacts of LLMs on society, human rights, and democratic institutions [29–33]. We shed light on the biases that shape the narratives these models tell about freedom of the press in their home countries and other nations. We show that these biases and misalignments are not random or uniform. Most importantly, we highlight the shortcomings of language models in accurately depicting human rights. We do this not to discourage their use, but to argue in favor of proper regulation and governance of LLMs that uphold, rather than undermine, democratic values and human rights.

Our findings underscore the need for further research that sheds light on LLMs' biases and representations of democratic institutions, what their risks are, and their potential impacts on our systems of governance. How do LLMs portray other democratic institutions? Do they suffer from similar misalignments when describing the state of human rights and violations around the world? Can these misrepresentations be harnessed by nations with advanced AI capabilities to project soft power through algorithmically biased representations of global political realities? As these technologies become integral to information infrastructure, ensuring their alignment with democratic principles represents not merely a technical challenge but a fundamental requirement for preserving democratic governance in the digital age.

## 2 Results

We find that the six LLMs in our study consistently underestimate press freedom compared to human expert assessments. ChatGPT, for example, underrates press freedom in 97% of the 180 countries evaluated, with other models showing similarly high rates: Gemini (96%), Qwen (93%), Falcon (89%), Mistral (87%), and even the best-performing model in this specific task, DeepSeek, underrating 71% of countries. We call this systematic underestimation a negative generalized misalignment.

To quantify this misalignment, we conceptualize LLMs' generalized misalignment as the difference between the WPFI scores for a given country using Equation 1. We then statistically evaluate the general misalignment relative to expert WPFI evaluations for the six LLMs in our study across the 180 countries in our sample, using the regression specification in Equation 2 described in the Methods section. The six LLMs almost systematically underrate press freedom compared to human experts, rating countries below the WPFI benchmark far above 50% of the time, as shown by the negative, statistically significant coefficients in Table 1 in the column labeled 'General Misalignment'.

$$\text{Misalignment}_{LLM} = \text{LLM Score} - \text{WPFI Score} \quad (1)$$

Figure 2 further illustrates the systematic nature of this pattern by showing the distribution of LLM scores at the survey category level for the 180 countries considered. In the figure, the distribution of LLM scores is visibly shifted leftward relative to the WPFI distribution, with each model's median score falling strictly below the WPFI median. This type of misalignment is consistent across the six models, occurring regardless of model origin, size, or architecture.

| | *Dependent variable: Misalignment* | | | | | |
|---|---|---|---|---|---|---|
| | CHATGPT | GEMINI | FALCON | QWEN | MISTRAL | DEEPSEEK |
| Generalized Misalignment | -16.84*** | -16.75*** | -16.60*** | -16.04*** | -11.85*** | -5.580*** |
| | (0.776) | (0.833) | (0.909) | (0.850) | (0.777) | (0.777) |
| Country FE | Yes | | | | | |
| $R^2$ | 0.677 | | | | | |
| Observations. | 6300 | | | | | |
| N Countries | 180 | | | | | |

**Table 1** Generalized Misalignment Coefficients by LLM. This table shows the regression coefficients resulting from Equation 2 at the question level, capturing the generalized negative misalignment of LLMs relative to the WPFI benchmark.

We next examine whether LLM misalignment is uniformly distributed or follows systematic patterns across the 180 countries in our sample. We find that LLMs exhibit differential misalignment across countries - a pattern where LLM score deviations from the human benchmark systematically depend on a country's level of press freedom as measured by the WPFI score. This pattern pattern reveals that LLM deviations are not uniformly or randomly distributed across countries, but rather reflect systematic biases in how LLMs assess press freedom.

To statistically evaluate this type of misalignment at the country level, we estimate the regression individually for each LLM, excluding its home country as described in Equation 2 in the Methods section. When combining all country-level misalignment scores across LLMs, we estimate a combined coefficient of β = -0.251 as shown in Figure 3, which confirms a consistent trend across LLMs.

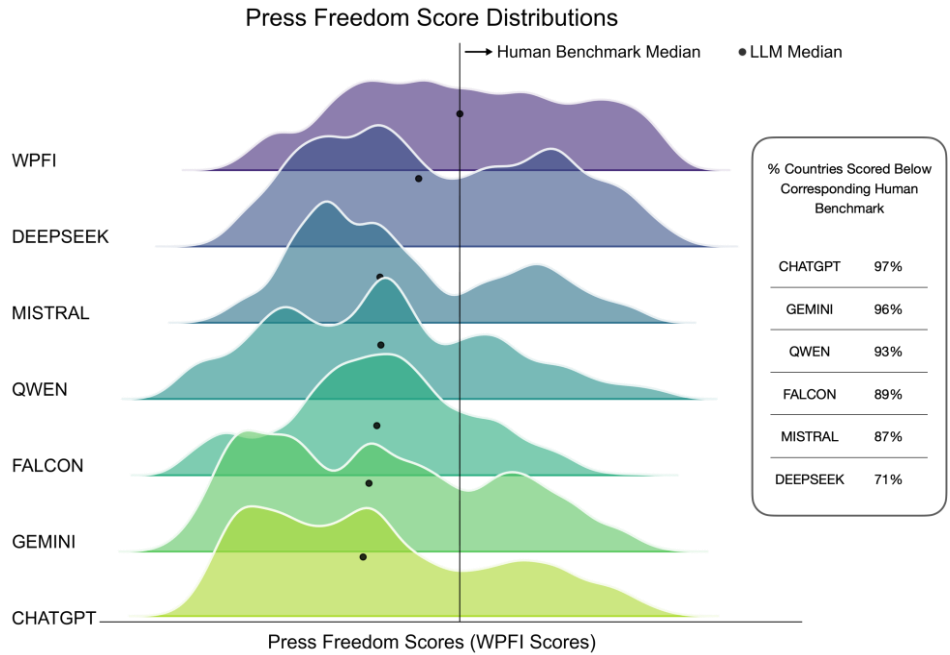

**Fig. 2 Distributions of press freedom scores across 180 countries from human experts and six LLMs.** This figure presents the distribution of press freedom scores for the 180 countries in our study. The top distribution (dark purple) corresponds to expert scores from the World Press Freedom Index (WPFI), followed by the distributions of scores generated by the six LLMs in our sample. The vertical black line indicates the WPFI median score, while the black dots mark the median of each distribution. All six LLM distributions are visibly shifted to the left of the WPFI benchmark, and their medians fall below the WPFI median, indicating a consistent pattern of more negative assessments by the language models.

The negative, statistically significant coefficients shown in Table 2 provide evidence that LLMs are more critical of countries that are rated as freer by human experts. Individually, Falcon shows the strongest differential misalignment, followed by Mistral and Qwen. Gemini and ChatGPT exhibit more moderate instances of this misalignment, while DeepSeek shows the least, though still statistically significant, differential misalignment. These findings remain robust when including each LLM's home country, but we find that estimating without it provides a better estimation due to the home bias these models exhibit.

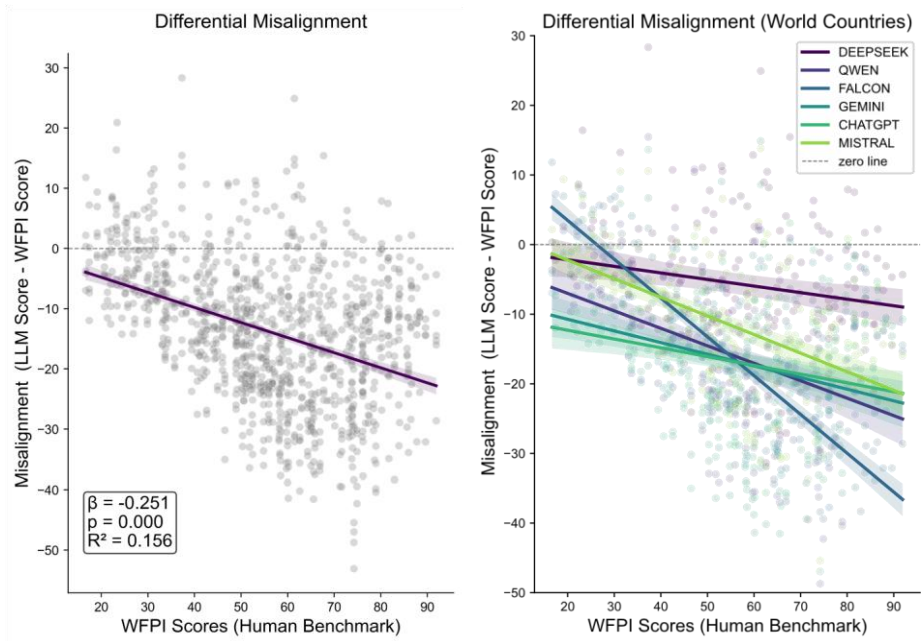

**Fig. 3 Differential misalignment in LLM press freedom evaluations.** This figure illustrates the differential negativity exhibited by the six LLMs in our study. Slope steepness indicates the extent of differential misalignment: flatter slopes suggest weaker differential misalignment, while steeper slopes indicate stronger effects. The left panel shows the overall trend when pooling all LLM scores. The right panel displays differential misalignment by individual LLM. This analysis excludes the four home countries of the LLMs studied. The line plots show that Falcon (blue) exhibits the strongest differential misalignment and DeepSeek (purple) the weakest.

These findings are also illustrated in Figure 3. The left panel displays the pooled regression results, while the right panel shows regression lines for each individual LLM. In this figure, a negative slope shows the negative relationship between the deviations between the LLM scores and the WPFI scores and the WPFI scores themselves. A flatter slope indicates a smaller degree of differential misalignment. Notably, the right panel reveals an interesting trend: the LLMs tend to agree more closely when evaluating countries with lower WPFI scores—typically more autocratic regimes, as shown by their regression lines clustering on the left side of the x-axis. In contrast, the models diverge significantly in their evaluations of countries with higher levels of press freedom, represented on the right side of the x-axis.

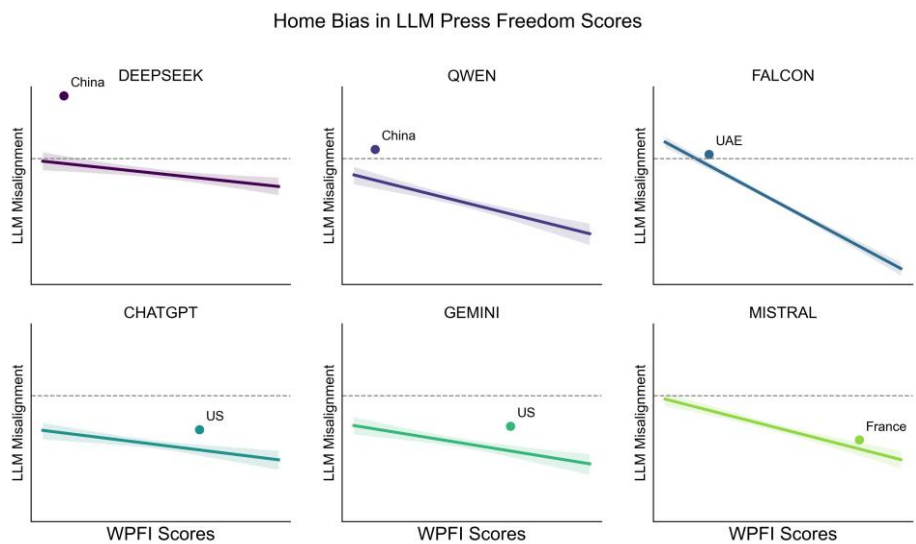

**Fig. 4 Home bias in LLM evaluations of Press Freedom.** This bar plot displays the home bias for each of the six LLMs in our study. Five of the six LLMs show a positive home bias. DeepSeek and Qwen exhibit the strongest biases, followed by Falcon, Gemini and ChatGPT. Mistral is the only LLM to display a negative home bias, rating France's press freedom below both the WPFI score and its own scoring baseline. The gray bars show the models' overall negativity towards the other 179 countries they were asked to evaluate.

Finally, we find that five of the six LLMs exhibit significant positive home bias, systematically rating press freedom in their home countries more favorably than expected by their general and differential misalignment. DeepSeek exhibits the strongest positive home bias, with a coefficient of 11.49, rating China nearly 20 points higher (on the 100-point scale) than the WPFI benchmark when examining raw score differences. Qwen shows similarly pronounced bias toward China with a coefficient of 9.2. Among Western models, Gemini exhibits substantial home bias for the US (coefficient = 9.53), followed by GPT-4 (coefficient = 8.06, also for the US). Falcon shows markedly lower home bias toward the UAE (coefficient = 1.9), while Mistral shows the weakest evidence of home bias for France (coefficient = 2.29) significant only at p ¡ 0.1 compared to p ¡ 0.01 for all others.

|  | DEEPSEEK | QWEN | FALCON | GPT | GEMINI | MISTRAL |
|---|---|---|---|---|---|---|
| Intercept | 4.445*** | 4.298** | 21.12*** | -6.022*** | 1.684 | 6.192*** |
|  | (1.531) | (1.784) | (1.362) | (1.487) | (1.700) | (1.321) |
| LLM Scores | -0.177*** | -0.359*** | -0.674*** | -0.192*** | -0.329*** | -0.320*** |
|  | (0.0256) | (0.0354) | (0.0265) | (0.0301) | (0.0345) | (0.0264) |
| $R^2$ | 0.0705 | 0.176 | 0.445 | 0.0719 | 0.140 | 0.200 |
| Observations | 895 | 895 | 895 | 895 | 895 | 895 |

\* $p < 0.1$, \*\* $p < 0.05$, \*\*\* $p < 0.01$

**Table 2 Differential misalignment of LLM models compared to WPFI Scores.** The table presents regression coefficients from separate models for each LLM, where the dependent variable is the difference between LLM and human expert scores. Negative coefficients for LLM Scores show that LLM misalignment is not random or uniform. It is associated with the values of press freedom themselves. Standard errors are reported in parentheses and the statistical significance is shown with one to three stars. This analysis excludes the four home countries of the LLMs, taking the number of observations from 180 to 176, to avoid mixing the effect of differential misalignment with home bias.

This home bias represents a specific case of differential misalignment, as models systematically favor their countries of origin while applying different standards elsewhere. Notably, this creates cross-national differences in LLMs evaluations of press freedom: while Chinese models rate China positively, both Falcon and Gemini have a negative bias toward China (Table in the Appendix), illustrating how geopolitical perspectives may be embedded in these AI systems' evaluative frameworks.

## 3  Discussion

Our analysis of six widely used LLMs across 180 countries reveals systematic distortions in their evaluation of press freedom. These distortions lead to a generalized underestimation of press freedom by LLMs, a differential evaluation that disproportionately penalizes freer societies, and a home bias favoring the models' home countries.

We believe that one of the mechanisms that leads to the differential misalignment we observe is rooted in asymmetries in information availability in the models' training data. Particularly, we refer to what some authors have called the democratic dilemma [34], whereby in societies with greater press freedom, journalists and citizens can freely critique government policies and document press freedom violations, generating substantial negative coverage that becomes overrepresented in the training data. Empirical evidence supports this mechanism: negative news has been shown to have higher salience, broader dissemination, and increased social media sharing compared to positive news [13, 35]. Conversely, authoritarian regimes suppress critical reporting about their press freedom or lack thereof, biasing the training data due to the absence of such information. We hypothesize the positive home bias we observe in five of the six LLMs is due to biases in the training data and the RLHF alignment process. We also believe this misalignment to be closely related to the in-group bias that has been found in LLMs. Together, these findings reveal the potential that LLMs have to distort the public's perception and trust of the democratic institutions they enjoy and those enjoyed by others. These distortions may downplay the restrictions imposed by authoritarian regimes on the press and foment distrust in already fragile democratic institutions. When people believe the media is freer than it is, they may not recognize when press freedoms are being systematically dismantled. Our insights also question the reliability of LLMs as tools for global human and civic rights assessments, underscoring the importance of honest and rigorous journalistic work. Another implication of the positive home bias is that LLMs may enable wealthier nations, or those with the capacity to build and deploy these models globally, to project soft power on the world stage. As LLMs increasingly mediate public understanding of political realities, their systematic misrepresentation of press freedoms and potentially other civic and human rights raises concerns about their role in shaping individuals' perceptions of different political regimes and geopolitical narratives.

Our study has a few limitations. First, our analysis relies on a single benchmark for press freedom: the WPFI. While a trusted non-governmental institution develops the WPFI, it represents only one framework for evaluating press freedom. Thus, there is potential for our findings not to generalize if researchers used a different instrument to measure press freedom. Second, we did not systematically try different variations of our prompts Because we wanted to mimic the closest way a regular user might prompt the LLM. LLM outputs can be sensitive to prompt design, so different formulations or additional context might yield different scores. Our findings highlight the urgent need for systematic evaluation of LLMs' representation of democratic institutions beyond press freedom. We propose that researchers probe into LLMs' representations of other democratic institutions and their representations of human rights. Further, we believe that measuring downstream effects on public opinion and policy preferences will quantify the real-world consequences of these biases for democratic governance. As these technologies become embedded in global information infrastructure, their alignment with democratic principles becomes not just a technical challenge but a fundamental requirement for preserving democratic societies in the digital age.

## 4 Methods

Three criteria guided our selection of LLMs: 1) the country of origin of the developing companies to ensure broad geographical representation, and 2 model capabilities to ensure that we surveyed the most capable question-answering model each company offers, and 3) how widely used the models are. Accordingly, we selected the following six models for our study: GPT-4.0 and Gemini-1.5, representing the United States; DeepSeek-Chat and Qwen25-32B, representing China; Mistral-Large-2411, representing France; and Falcon-180B, representing the United Arab Emirates.

We used API calls to collect responses for all models except for Falcon-180B, which did not have an accessible API between February and January 2025, the time we conducted our study. Instead, we used the AI71 platform1, to collect answers for Falcon.

### 4.1 Text Generation

To evaluate how LLMs assess press freedom worldwide, we created a set of prompts based on the World Press Freedom survey developed by the non-profit, non-governmental organization Reporters Without Borders (RWB) [36].

As its name suggests, the survey aims to assess the state of press freedom annually in 180 countries. It consists of multiple-choice questions spanning five categories of press freedom: political, social, economic, safety, and legal. This survey underpins the World Press Freedom Index (WPFI), RWB's expert assessment of press freedom. The experts in RWB's panel include journalists, researchers, and regional specialists with local knowledge. These experts draw on field reporting, legal frameworks, and institutional sources to evaluate the state of press freedom worldwide. RWB annually publishes WPFI scores at the category level for all participating nations.

---

[1] https://ai71.ai/

To adapt the survey or our study, we modified each original question by appending a string instructing the LLM to respond with reference to a specific country (see Appendix for details). We did not provide the LLMs with additional context beyond the modified survey question. We used a temperature setting of 0.8 for all models. Temperature is an LLM parameter that controls the randomness of responses: lower values (closer to 0) produce more deterministic and consistent outputs, while higher values introduce greater variation and creativity in the generated text.

Because LLMs are known to be sensitive to the order of multiple-choice answers, we conducted three survey waves for all LLMs. The first wave spanned 180 countries and used the original answer choice order. The subsequent two surveys randomized the order of the possible answer choices for each question. These two subsequent surveys included only 22 countries, selected using a stratified sample based on GDP and WPFI scores (see Appendix for details). Each LLM was prompted 21,060 times (117 questions × 180 countries) with the original answer order and 5,148 times (2 randomizations × 22 countries × 117 questions) with a randomized answer order. Shuffling answer order did not significantly alter our results.

## 4.2 Press Freedom Scores

RWB scores each country on a scale from 1 to 100, with higher values indicating greater press freedom. When we started our study, the most recent edition of the WPFI was the 2024 release. At the time of writing, the 2025 version was released, but it was not included due to time constraints. We use these scores as the human benchmark for our study and refer to them as WPFI scores throughout the paper.

To compute the LLM-based survey scores, we follow RWB's methodology and assign a score between 1 and 100 to each country based on the LLM's multiple-choice responses. We refer to these as LLM scores throughout the paper. Among the five survey categories, safety is the only one for which our approach does not fully replicate RWB's methodology: specifically, we do not include outside information on physical attacks and attempts on the lives of journalists, which are included in the original index from RWB's Barometer data [37]. We base our scores exclusively on the LLM responses to the survey.

We do not expect this change in how the safety scores are calculated to affect our results significantly. Most of the information in RWB's Barometer describes the challenges and dangers faced by reporters; therefore, incorporating this information into the LLM scores for a final assessment of reporter safety would likely further lower the press freedom scores we obtain from our surveys and increase the magnitude of our results. However, to test the robustness of our findings, we run our analysis excluding this category and conclude that our results remain unchanged, as shown in the Appendix.

Furthermore, we find that the WPFI scores are positively correlated with media trust percentages reported by Reuters Institute and Oxford University in their Digital News Report from 2024. For more details see the Appendix.

## 4.3 Differential Misalignment and Home Bias

We estimated the model described in Equation 2 in two ways: First, we pooled observations across all six LLMs to capture the common trend across all language models in our study. Then, we regressed each LLM separately to examine LLM-specific differential misalignment. We estimated the model in Equation 2 for each LLM, including and excluding their home countries, to avoid potential contamination of these results with each LLM's home bias. Here, we report the results excluding the LLMs' home country and include the results in the Appendix.

$$\text{Misalignment}_{\text{LLM}} = \alpha + \beta \text{WPFI Scores} + \epsilon \qquad (2)$$

We also investigate a specific case of LLMs' differential misalignment, one related to in-group bias and driven by the language model's home country. In particular, we wanted to test if LLMs would depict the freedom of the press in their country of origin as more favorable relative to the human benchmark and other language models in our sample. We assign each LLM a home country based on the location of the company that developed it. To formally test for home bias, we estimate the regression model shown in Equation Y. We estimate this regression for all three survey rounds to ensure the robustness of our results.

$$\text{Score}_{i,j} = \alpha \text{Home}_i + \beta \text{LLM}_i + \gamma(\text{Home}_i \text{LLM}_i) + \text{CountryFE} + \epsilon_{i,j} \qquad (3)$$

In Equation 3, $\text{Home}_i$ is a dummy variable equal to one if the country is the LLM's home country. $\text{LLM}_i$ is another dummy equal to one if the score is generated by an LLM, and zero if the score comes from the WPFI. The interaction between an LLM's home country and the dummy indicating that the same LLM generated the score, captured by γ, represents the LLM's home bias. If the LLM systematically assigns a higher score to its own country compared to other countries and other LLMs, it would exhibit a positive home bias. In contrast, a negative γ would indicate a negative home bias. That is, an LLM that systematically underrates press freedom in its home country. Country FE control for country-specific unobserved heterogeneity.

### 4.4 WPFI Scores and Media Trust

To validate our independent variable and understand whether the WPFI is associated with media trust, we used public data from the Reuters Institute Digital News Report, published annually by the Reuters Institute and the Oxford University Internet Institute. Specifically, we used their Media Trust variable, which measures the proportion of people who trust "most news most of the time" [38] across 47 countries. This data is reported as a percentage.

When we regress media trust data on WPFI, we find that over the last three years, the relationship between WPFI and trust data is positive and statistically significant.

We used data from 2023 onward because Reporters Without Borders changed their methodology in 2023, creating a break in the time series. We therefore focus on how trust correlates with the current methodology for measuring press freedom. We removed extreme outliers from the yearly regressions using Cook's D. Nigeria and Kenya were consistently detected as outliers across all three years because, despite having low press freedom scores, they show very high percentages of people who trust the media. In 2024, Finland and Hong Kong were also detected as outliers and were therefore removed. The regression coefficients and Cook's D values for 2024 are shown in Table 3 and Table 4, respectively.

| Dependent variable: % people who trust the media | | | |
|---|---|---|---|
| | 2022 | 2023 | 2024 |
| Intercept | 24.40*** | 21.32*** | 20.37*** |
| | (6.78) | (7.29) | (7.36) |
| WPFI Score | 0.251** | 0.259** | 0.210** |
| | (0.098) | (0.102) | (0.104) |
| Observations | 43 | 43 | 41 |
| R-squared | 0.139 | 0.136 | 0.089 |

**Table 3** WPFI Scores and Trust in the Media. This table shows that WPFI Scores and media trust are statistically correlated between 2022 and 2024.

| Country | WPFI Score | Media Trust (%) | Cook's Distance |
| --- | --- | --- | --- |
| Finland | 86.55 | 69.0 | 0.170 |
| Hong Kong | 43.06 | 55.0 | 0.120 |
| Kenya | 53.22 | 64.0 | 0.119 |
| Nigeria | 51.03 | 61.0 | 0.111 |

**Table 4** Cooks Distance to detect Influential observations. In 2024, Finaland, Hong king, Kenya and Nigeria were removed because they all had values greater than Cook's D (Cook's D > 0.089).

Our analysis of six widely used LLMs across 180 countries reveals three systematic misalignments in their evaluation of press freedom. These distortions lead to a generalized underestimation of press freedom by LLMs, a differential misalignment that disproportionately penalizes freer societies, and a home bias favoring the models' home countries.

We believe that one of the mechanisms that leads to the differential misalignment we observe is rooted in asymmetries in information availability in the models' training data. Particularly, we refer to what some authors have called the democratic dilemma [34], whereby in societies with greater press freedom, journalists and citizens can freely critique government policies and document press freedom violations, generating substantial negative coverage that becomes overrepresented in the training data. Empirical evidence supports this mechanism: negative news has been shown to have higher salience, broader dissemination, and increased social media sharing compared to positive news [13, 35].

Conversely, authoritarian regimes suppress critical reporting about their press freedom or lack thereof, biasing the training data due to the absence of such information. We hypothesize that the positive home bias we observe in five of the six LLMs is connected to in-group bias that has been previously found in LLMs. If this is the case, then alignment techniques such as instruction fine-tuning and preference-tuning could be effective at reducing this type of bias, as they were suggested to be effective at addressing social identity bias [39].

Together, these findings reveal the potential that LLMs have to distort the public's perception and trust in the press and its independence. Given that press freedom as measure by the WPFI correlates with citizen trust (as shown in our validation in the Appendix), LLM misalignments may have downstream effects on public confidence in the media. These distortions may downplay the restrictions imposed by authoritarian regimes on the press and foment distrust in already fragile democratic institutions. When people believe the media is less free than it is, they may not recognize when press freedoms are being systematically dismantled. Our insights also question the reliability of LLMs as tools for global human and civic rights assessments, underscoring the importance of honest and rigorous journalistic work.

Another implication of the positive home bias is that LLMs may enable wealthier nations, or those with the capacity to build and deploy these models globally, to project soft power on the world stage. As LLMs increasingly mediate public understanding of political realities, their systematic misrepresentation of press freedoms and potentially other civic and human rights raises concerns about their role in shaping individuals' perceptions of different political regimes and geopolitical narratives.

Our study has a few limitations. First, our analysis relies on a single benchmark for press freedom: the WPFI. While a trusted non-governmental institution develops the WPFI, it represents only one framework for evaluating press freedom. Second, we did not systematically try different variations of our prompts because we wanted to mimic the closest way a regular user might prompt the LLM. LLM outputs can be sensitive to prompt design, so different formulations or additional context might yield different scores.

Our findings highlight the urgent need for systematic evaluation of LLMs' representation of democratic institutions beyond press freedom. We propose that researchers probe into LLMs' representations of other democratic institutions and their representations of human rights. Furthermore, we believe that measuring downstream effects on public opinion and policy preferences will quantify the real-world consequences of these biases for democratic governance. As these technologies become embedded in global information infrastructure, their alignment with democratic principles becomes not just a technical challenge but a fundamental requirement for preserving democratic societies in the digital age.

## Declarations

The authors have no competing interests. Data and code pipeline to reproduce results will be made public on github after publication. In the meantime, it will be available upon request.

## References

[1] Mishler, W. & Rose, R. What are the origins of political trust? testing institutional and cultural theories in post-communist societies. *Comparative political studies* **34**, 30–62 (2001).

[2] Tsfati, Y. & Cohen, J. Democratic consequences of hostile media perceptions: The case of gaza settlers. *Harvard International Journal of Press/Politics* **10**, 28–51 (2005).

[3] Ladd, J. M. The role of media distrust in partisan voting. *Political Behavior* **32**, 567–585 (2010).

[4] Hooghe, M. Trust and elections. em uslaner (2018).

[5] Hanitzsch, T., Van Dalen, A. & Steindl, N. Caught in the nexus: A comparative and longitudinal analysis of public trust in the press. *The international journal of press/politics* **23**, 3–23 (2018).

[6] Tsfati, Y. & Ariely, G. Individual and contextual correlates of trust in media across 44 countries. *Communication research* **41**, 760–782 (2014).

[7] Guess, A. M., Barber´a, P., Munzert, S. & Yang, J. The consequences of online partisan media. *Proceedings of the National Academy of Sciences* **118**, e2013464118 (2021).

[8] Newman, N. & Fletcher, R. *Bias, bullshit and lies: Audience perspectives on low trust in the media* (Reuters Institute for the Study of Journalism, 2017).

[9] Kalogeropoulos, A., Suiter, J., Udris, L. & Eisenegger, M. News media trust and news consumption: Factors related to trust in news in 35 countries. *International journal of communication* **13**, 22 (2019).

[10] Lorenz-Spreen, P., Oswald, L., Lewandowsky, S. & Hertwig, R. A systematic review of worldwide causal and correlational evidence on digital media and democracy. *Nature human behaviour* **7**, 74–101 (2023).

[11] Bak-Coleman, J. B. *et al.* Stewardship of global collective behavior. *Proceedings of the National Academy of Sciences* **118**, e2025764118 (2021).

[12] Cinelli, M., De Francisci Morales, G., Galeazzi, A., Quattrociocchi, W. & Starnini, M. The echo chamber effect on social media. *Proceedings of the national academy of sciences* **118**, e2023301118 (2021).

[13] Aral, S. *The hype machine: How social media disrupts our elections, our economy, and our health–and how we must adapt* (Crown Currency, 2021).

[14] Farrell, H., Gopnik, A., Shalizi, C. & Evans, J. Large ai models are cultural and social technologies. *Science* **387**, 1153–1156 (2025).

[15] Pacheco, A. G., Cavalini, A. & Comarela, G. Echoes of power: Investigating geopolitical bias in us and china large language models. *arXiv preprint arXiv:2503.16679* (2025).

[16] Schmidt, E. Innovation power: why technology will define the future of geopolitics. *Foreign Aff.* **102**, 38 (2023).

[17] Sun, Y. & Wang, T. Be friendly, not friends: How llm sycophancy shapes user trust. *arXiv preprint arXiv:2502.10844* (2025).

[18] Tao, Y., Viberg, O., Baker, R. S. & Kizilcec, R. F. Cultural bias and cultural alignment of large language models. *PNAS nexus* **3**, pgae346 (2024).

[19] Yang, K. *et al.* Unpacking Political Bias in Large Language Models: A Cross-Model Comparison on U.S. Politics (2025). URL http://arxiv.org/abs/2412.16746. ArXiv:2412.16746 [cs].

[20] Feng, S., Park, C. Y., Liu, Y. & Tsvetkov, Y. From pretraining data to language models to downstream tasks: Tracking the trails of political biases leading to unfair nlp models. *arXiv preprint arXiv:2305.08283* (2023).

[21] Rozado, D. The political biases of chatgpt. *Social Sciences* **12**, 148 (2023).

[22] Motoki, F., Pinho Neto, V. & Rodrigues, V. More human than human: measuring chatgpt political bias. *Public Choice* **198**, 3–23 (2024).

[23] Manvi, R., Khanna, S., Burke, M., Lobell, D. & Ermon, S. Large Language Models are Geographically Biased (2024). URL http://arxiv.org/abs/2402.02680. ArXiv:2402.02680 [cs].

[24] Sheng, E., Chang, K.-W., Natarajan, P. & Peng, N. The woman worked as a babysitter: On biases in language generation. *arXiv preprint arXiv:1909.01326* (2019).

[25] Bordia, S. & Bowman, S. R. Identifying and reducing gender bias in word-level language models. *arXiv preprint arXiv:1904.03035* (2019).

[26] Hu, T. *et al.* Generative language models exhibit social identity biases. *Nature Computational Science* **5**, 65–75 (2025).

[27] Durmus, E. *et al.* Towards measuring the representation of subjective global opinions in language models. *arXiv preprint arXiv:2306.16388* (2023).

[28] Zhou, D. & Zhang, Y. Political biases and inconsistencies in bilingual GPT models—the cases of the U.S. and China. *Scientific Reports* **14**, 25048 (2024). URL https://www.nature.com/articles/s41598-024-76395-w.

[29] An, J., Huang, D., Lin, C. & Tai, M. Measuring gender and racial biases in large language models: Intersectional evidence from automated resume evaluation. *PNAS nexus* **4**, pgaf089 (2025).

[30] Jiang, L., Zhu, G., Sun, J., Cao, J. & Wu, J. Exploring the occupational biases and stereotypes of chinese large language models. *Scientific Reports* **15**, 1–15 (2025).

[31] Salecha, A. *et al.* Large language models display human-like social desirability biases in big five personality surveys. *PNAS nexus* **3**, pgae533 (2024).

[32] Fang, X. *et al.* Bias of ai-generated content: an examination of news produced by large language models. *Scientific Reports* **14**, 5224 (2024).

[33] Briggs, M. & Cross, M. Generative ai: Threatening established human rights instruments at scale (2024).

[34] Heinze, E. & Freedman, R. Public awareness of human rights: distortions in the mass media. *The International Journal of Human Rights* **14**, 491–523 (2010).

[35] Watson, J., van der Linden, S., Watson, M. & Stillwell, D. Negative online news articles are shared more to social media. *Scientific Reports* **14**, 21592 (2024).

[36] Reporters Without Borders. World press freedom index (2024). URL https://rsf.org/en. Accessed: 2024-06-01.

[37] Reporters Without Borders. Press freedom barometer (2024). URL https://rsf.org/en/barometer. Accessed: 2024-06-01.

[38] Newman, N., Fletcher, R., Robertson, C. T., Ross Arguedas, A. & Nielsen, R. K. *Reuters Institute digital news report 2024* (Reuters Institute for the study of Journalism, 2024).

[39] Hu, T. *et al.* Generative language models exhibit social identity biases. *Nature Computational Science* **5**, 65–75 (2025). URL https://www.nature.com/articles/s43588-024-00741-1. Publisher: Nature Publishing Group.

[40] Pezeshkpour, P. & Hruschka, E. Large language models sensitivity to the order of options in multiple-choice questions. *arXiv preprint arXiv:2308.11483* (2023).

# Appendix A    Extended Data

## A.1    The World Press Freedom Index

The World Press Freedom Index, published annually by Reporters Without Borders (RWB), evaluates press freedom globally through a multiple-choice questionnaire designed to assess the capacity of journalists to operate independently of political, economic, legal, and social pressures. According to the official definition on their website[2], press freedom is defined as "...the ability of journalists as individuals and collectives to select, produce, and disseminate news in the public interest independent of political, economic, legal, and social interference and the absence of threats to their physical and mental safety."

The index assesses five key dimensions of press freedom: political context, legal framework, economic environment, sociocultural context, and safety. A panel of experts evaluates 180 countries with a qualitative study based on the responses to a questionnaire by journalists, researchers, academics, and human rights defenders. Only a fifteenth of each country's score is derived from a quantitative analysis, which tracks the number of abuses against journalists and media outlets in connection with their work. In this study, we only take into account the results from the survey questionnaire, that is the qualitative analysis.

There are 117 questions in the WPFI questionnaire distributed across the five domains: 33 in the political domain, 25 in the legal domain, 25 in the economic domain, 22 in the sociocultural domain, and 12 in the security domain. Each country receives a score ranging from 0 to 100 for each of these five categories and an overall score as the arithmetic mean of the five categories' scores. RSF does not disclose the scores for individual questions. As a result, we can only observe six observations for each country, taken from the latest version of the WPFI (2024): five category scores and an overall score. Each category contributes equally to the final score, and each question within a category carries the same weight.

The original questions in the survey are country-agnostic, for example:

> "What influence does the government have in the editorial board of media?"

---

[2] https://rsf.org/en/methodology-used-compiling-world-press-freedom-index-2024?year=2024& data type=general

Since we want to perform a multi-country analysis by providing the WPFI questionnaire as input to each Large Language Model (LLM), we adapted each question in the survey to make it country-specific. To do so, we added a country-specific prefix, "In country," where "country" changes accordingly to the country under analysis. For Uruguay, the adapted prompt becomes:

> "In Uruguay, what influence does the government have in the editorial board of media?"

The questionnaire also includes sub-questions presented in a grid format. For example:

> "What influence does economic power have over the editorial board of the following types of media? Public media / Private media"

We converted each grid item into a distinct prompt. This process resulted in two separate prompts:

1. "In Uruguay, what influence does economic power have over the editorial board on public media?"
2. "In Uruguay, what influence does economic power have over the editorial board on private media?"

For what concerns the answer modality, each question in the survey offers a limited set of multiple-choice answers, ranging from a maximum of five to a minimum of two options. The answers are pre-ordered by RWB, with the first option representing the best possible outcome and the last option representing the worst. We assigned a score to each response on a scale from 0 to 100, in line with RWB methodology. The best answer always receives a score of 100, while the worst answer receives a 0. When a question offers five possible answers, the scoring is distributed evenly: 100 for the 1st answer, 75 for the 2nd answer, 50 for the 3rd answer, 25 for the 4th and 0 for the 5th. For questions with only two answer choices, the best answer receives a score of 100, and the worst answer receives a score of 0.

**Prompt 1**:

> System message: You are a concise assistant. Always answer with only the letter corresponding to the correct answer choice, without any explanations or additional text.
>
> User message: Question: In Cameroon, is the broadcast licensing process transparent?
>
> Choices:
> A  - Yes, completely
> B  - Yes, somewhat
> C  - No opinion
> D  - Not veryE - No, not at all Answer:

## A.2 Regression Robustness with Randomized Order

Since the input prompts provided to the various LLMs consist of multiple-choice questionnaires, one potential concern is that LLMs' responses may be sensitive to the order in which the answer options are presented. Recent studies have highlighted that LLMs can indeed be influenced by positional bias in multiple-choice settings [40]. Another relevant concern relates to the temperature parameter: setting it to 0.8, as done in the main analysis, may introduce additional randomness in model responses compared to lower temperature settings. To address these issues, we conducted a robustness check using a subset of 22 countries, selected to ensure a balanced representation across levels of GDP and press freedom (as measured by WPFI scores).

The subset also includes the four countries of origin for the LLMs analyzed: China, the UAE, the United States, and France. For each country, we ran the analysis twice, randomizing the order of answer options in each instance to test for positional sensitivity.

**Table A1**: Selected Regression Results — Randomized and Open-Ended Models

|                  | (1) Randomization 1 | (2) Randomization 2 | (3) Open-ended |
|------------------|---------------------|---------------------|----------------|
| DEEPSEEK         | -10.18***           | -9.197***           | -5.580***      |
|                  | (2.780)             | (2.646)             | (0.777)        |
| FALCON           | -14.79***           | -14.54***           | -16.60***      |
|                  | (3.147)             | (3.323)             | (0.909)        |
| GEMINI           | -23.91***           | -23.94***           | -16.75***      |
|                  | (3.014)             | (3.100)             | (0.833)        |
| GPT              | -18.93***           | -18.67***           | -16.84***      |
|                  | (2.474)             | (2.458)             | (0.776)        |
| MISTRAL          | -16.64***           | -16.48***           | -11.85***      |
|                  | (3.026)             | (2.991)             | (0.777)        |
| QWEN             | -11.56***           | -11.03***           | -16.04***      |
|                  | (3.374)             | (3.170)             | (0.850)        |
| DEEPSEEK × China | 17.99***            | 15.34***            | 26.51***       |
| | (2.780) | (2.646) | (0.777) |
| FALCON × UAE | 17.34*** | 20.46*** | 18.02*** |
| | (3.147) | (3.323) | (0.909) |
| GEMINI × US | 11.86*** | 9.059*** | 6.514*** |
| | (3.014) | (3.100) | (0.833) |
| GPT × US | 6.679** | 9.670*** | 5.471*** |
| | (2.474) | (2.458) | (0.776) |
| MISTRAL × France | -3.146 | -3.437 | -2.907*** |
| | (3.026) | (2.991) | (0.777) |
| WPFI Scores | 0.570*** | 0.495*** | 0.479*** |
| | (0.0710) | (0.0822) | (0.0273) |
| R-squared | 0.623 | 0.606 | 0.677 |
| Observations | 770 | 770 | 6300 |
| N Countries | 22 | 22 | 180 |

Standard errors in parentheses. * $p < 0.1$, ** $p < 0.05$, *** $p < 0.01$

Table A1 reports the results across three specifications for the 22 countries: Column 1 presents the results using the original prompt order, while Columns 2 and 3 present results using randomly shuffled answer choices. Across all specifications, the findings are consistent with our main results. Specifically, all LLMs continue to systematically underestimate press freedom, reaffirming the presence of a global negative misalignment relative to WPFI scores.

Regarding the relationship between press freedom and model evaluation, we again observe that countries with lower press freedom (e.g, China and UAE) tend to receive more positive misalignment scores, suggesting that LLMs are more likely to suppress negativity in these contexts. This reinforces our central argument that LLMs mirror the availability of critical content in their training data, underestimating censorship in countries where such content is scarce.

As for home bias, the results remain consistent. For China, Eastern models—particularly Qwen and DeepSeek—continue to display strong positive biases, larger than those of Western models. Similarly, for the UAE, Falcon and Qwen again exhibit the most pronounced positive biases across specifications. For France, all models remain largely neutral, with the exception of Falcon, which continues to show a negative bias, consistent with the main analysis. Finally, for the United States, Western models—along with Qwen—show a positive bias, while no significant home bias is observed for DeepSeek or Falcon.

In sum, even with a higher temperature setting (0.8) and randomized prompt order, the results remain robust. The key patterns identified in the main analysis—systematic underestimation of press freedom, the influence of press freedom on misalignment, and the presence of home bias—are replicated, confirming the stability and reliability of our findings.

The following tables are robustness checks for differential misalignment between LLMs.

|  | DEEPSEEK | QWEN | FALCON | GPT | GEMINI | MISTRAL |
|---|---|---|---|---|---|---|
| Intercept | -1.059 | -0.134 | 4.224*** | -3.907*** | -8.031*** | -2.998** |
|  | (1.508) | (1.368) | (1.456) | (1.308) | (1.545) | (1.269) |
| LLM Scores | -0.219*** | -0.207*** | -0.280*** | -0.214*** | -0.261*** | -0.207*** |
|  | (0.0268) | (0.0245) | (0.0268) | (0.0262) | (0.0336) | (0.0237) |
| $R^2$ | 0.125 | 0.111 | 0.148 | 0.116 | 0.144 | 0.122 |
| Observations | 895 | 895 | 895 | 895 | 895 | 895 |

* $p < 0.1$, ** $p < 0.05$, *** $p < 0.01$

**Table A2 Differential misalignment of LLM models compared to WPFI Scores for the open-ended answers.** The table presents regression coefficients from separate models for each LLM, where the dependent variable is the difference between LLM and human expert scores. Negative coefficients for LLM Scores show that LLM misalignment is associated with the values of press freedom themselves. Standard errors are reported in parentheses and the statistical significance is shown with one to three stars.

|  | DEEPSEEK | QWEN | FALCON | GPT | GEMINI | MISTRAL |
|---|---|---|---|---|---|---|
| Intercept | 8.524* | 21.41*** | 25.20*** | 2.204 | 6.523 | 9.467** |
|  | (4.681) | (5.637) | (4.204) | (3.253) | (3.790) | (3.569) |
| LLM Scores | -0.312*** | -0.558*** | -0.715*** | -0.371*** | -0.539*** | -0.456*** |
|  | (0.0862) | (0.109) | (0.0735) | (0.0773) | (0.0877) | (0.0832) |
| $R^2$ | 0.201 | 0.320 | 0.482 | 0.274 | 0.441 | 0.418 |
| Observations | 105 | 105 | 105 | 105 | 105 | 105 |

* $p < 0.1$, ** $p < 0.05$, *** $p < 0.01$

**Table A3 Differential misalignment of LLM models compared to WPFI Scores for the randomized sample.** The table presents regression coefficients from separate models for each LLM, where the dependent variable is the difference between LLM and human expert scores on a subset of observations. Negative coefficients for LLM Scores indicate the association of LLM misalignment with press freedom values. Standard errors are in parentheses, and significance levels are indicated by stars.